\documentclass[12pt]{article}

\usepackage{graphicx} \usepackage{amssymb, amsmath}
\def\ave#1{\left\langle #1 \right\rangle} 

\begin{document}

\title{\bf Cosmic Shear from STIS Pure Parallels \\ \Large II Analysis}

\author{H.~H{\"a}mmerle$^{1,2}$,
 \and J.-M.~Miralles$^{3}$,
 \and P.~Schneider$^{1,2}$,
 \and T.~Erben$^{2,4,5}$,
 \and R.~A.~E.~Fosbury$^{3}$, 
 \and W.~Freudling$^{3}$, 
 \and N.~Pirzkal$^{3}$, 
 \and S.~D.~M.~White$^{2}$ 
} 

\date{\small
 $^{1}$Institut f{\"u}r Astrophysik und Extraterrestrische
 Forschung der Universit{\"a}t Bonn, Auf dem H{\"u}gel 71, D-53121
 Bonn, Germany \\
 $^{2}$Max-Planck-Institut f{\"u}r Astrophysik, Karl-Schwarzschild
 Str. 1, D-85748 Garching, Germany \\
 $^{3}$ST-ECF, Karl-Schwarzschild Str. 2,
 D-85748 Garching, Germany\\
 $^{4}$Institut d'Astrophysique de Paris, 98bis Boulevard Arago,
 F-75014 Paris, France \\
 $^{5}$Observatoire de Paris, DEMIRM 61, Avenue de l'Observatoire,
 F-75014 Paris, France  }

\maketitle

\section{Cosmic Shear}

The measurement of cosmic shear requires deep imaging (to beat the
noise due to the galaxies' intrinsic ellipticity distribution) with
high image quality (because every non-corrected PSF anisotropy mimics
cosmic shear) on many lines of sight to sample the statistics of
large-scale structure. The STIS camera on-board HST has a very good 
performance in that respect, which we demonstrate here, by using
archival data from the STIS pure parallel program between June 1997
and October 1998. The data reduction and catalog production are
described in the poster \emph{Detection of Cosmic Shear From STIS
Parallel Archive Data: Data Analysis} presented here and the paper
Pirzkal et al. (2001). 

\section{STIS: PSF correction}

\begin{figure}[!b]  
 \center{ 
 \includegraphics[width=7cm]{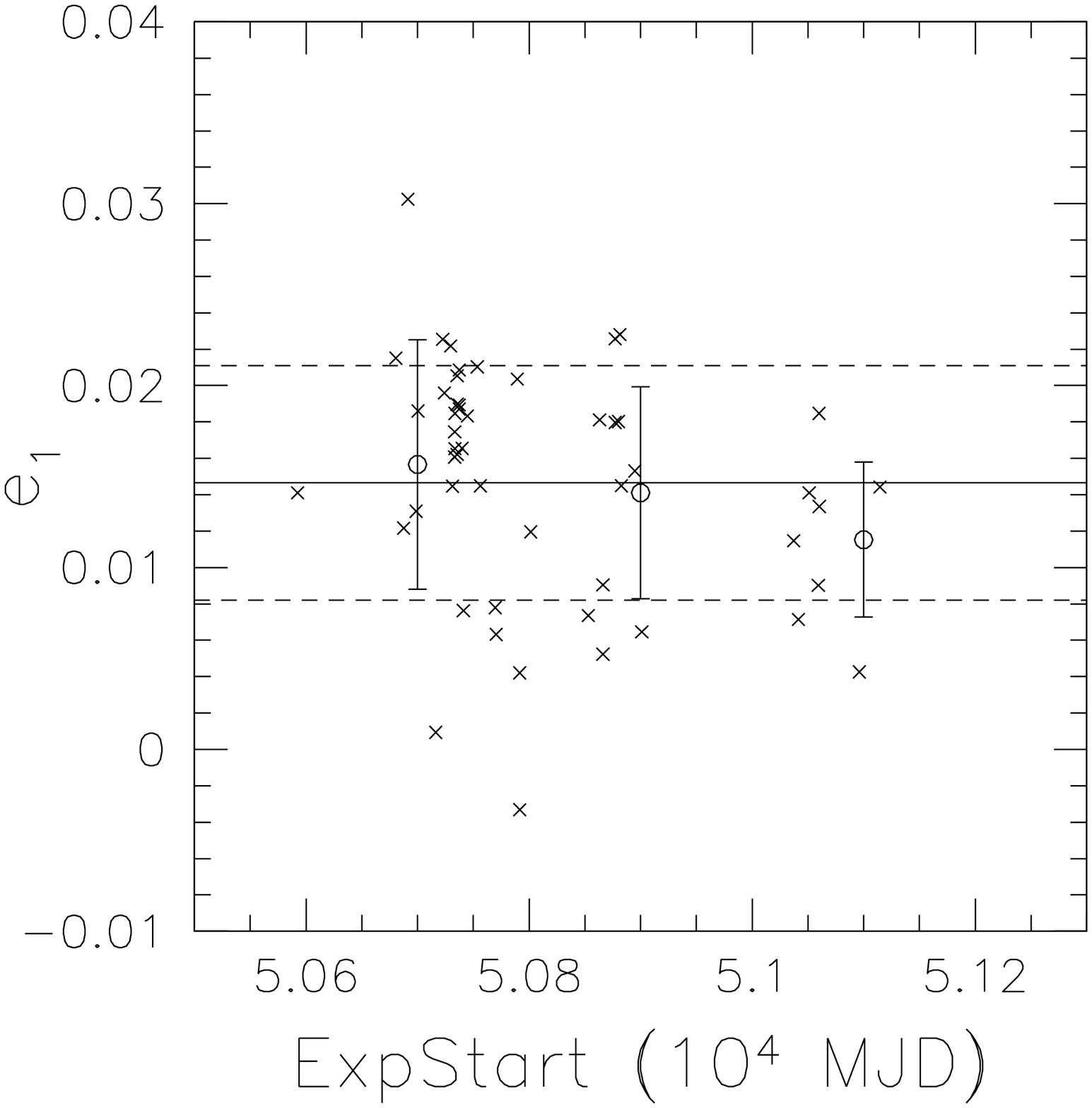} \hspace{1.5cm}
 \includegraphics[width=7cm]{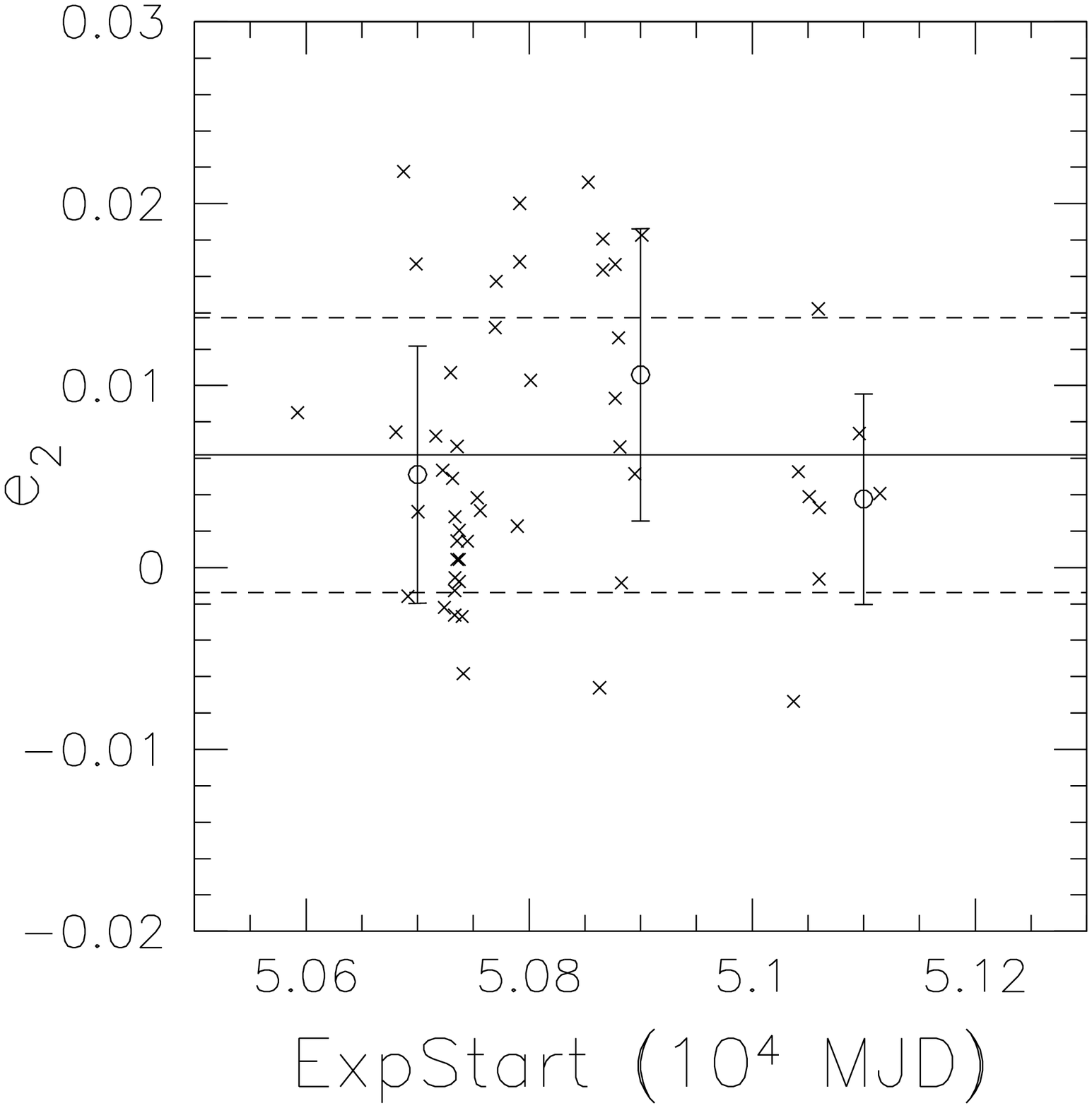}}
 \caption{Mean of the ellipticity $e_1$ (left) and $e_2$ (right) of
 the star fields vs. exposure start time (Modified Julian Date). Both
 components can be approximated by a constant over the time period
 covered. The staight lines show the mean over all the fields, the
 dashed lines are the corresponding $1\sigma$ errors. The circles
 denote the mean over stars in bins between $5.06\times 10^4$, 
 $5.08\times 10^4$, $5.10\times 10^4$, $5.12\times 10^4$ (MJD) with 
 the corresponding 1$\sigma$ error bars.  }
\label{fig:aniexp}
\end{figure}

The expected distortion of galaxy images by cosmic shear on the 
angular scale of STIS ($\sim 50^{\prime\prime}$) is a few percent, therefore
the PSF anisotropy has to be 
understood and controlled to an accuracy of better than 1\%. In
Fig.~\ref{fig:aniexp} we show that the mean ellipticity of stars are
small and can be considered constant at the one sigma level. 

\begin{figure}[!h]                             
\begin{minipage}{11.5cm}                      
 \includegraphics[width=11cm]{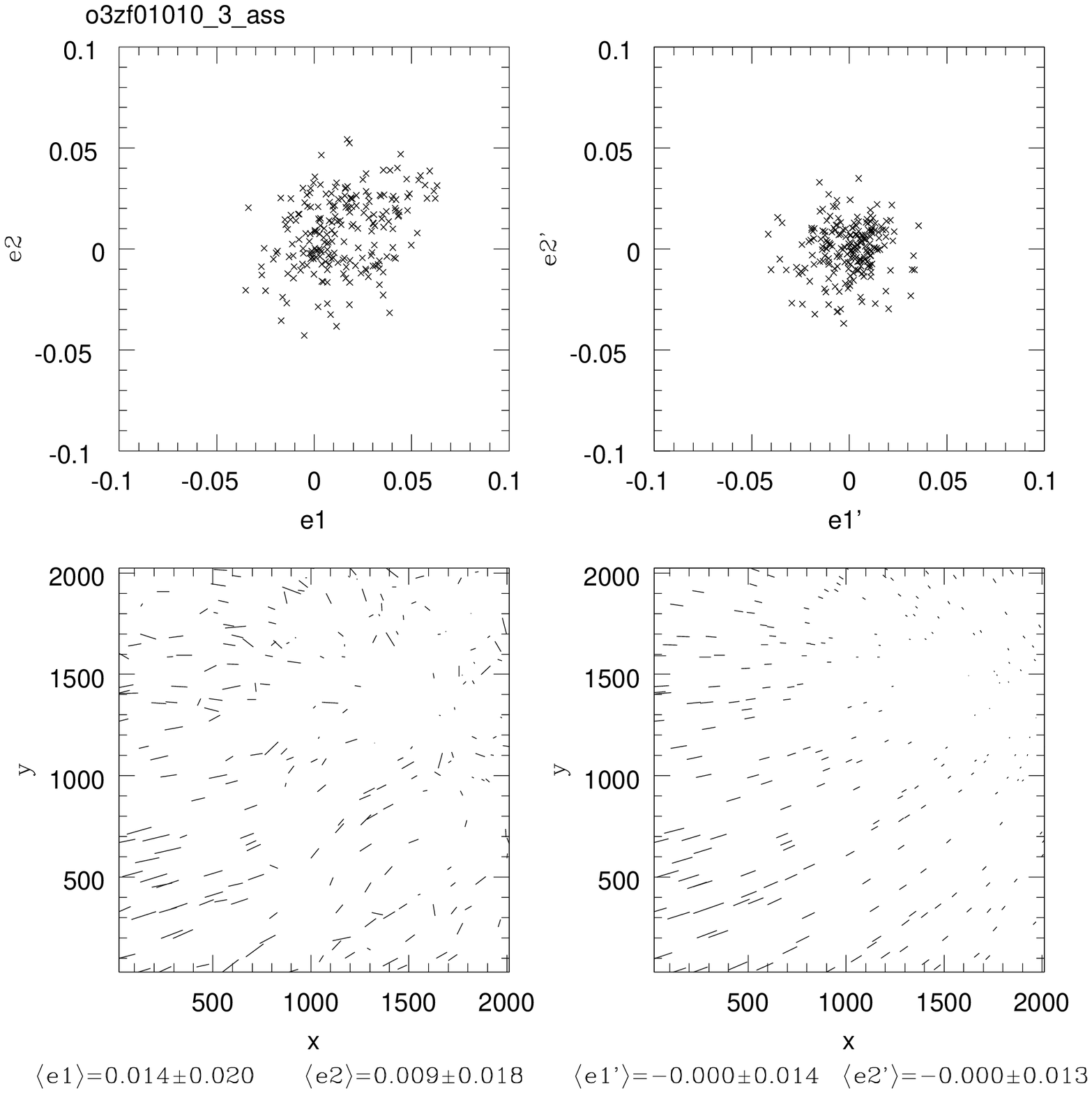}\\
 \mbox{}\hspace{0.26cm}\includegraphics[width=5cm]{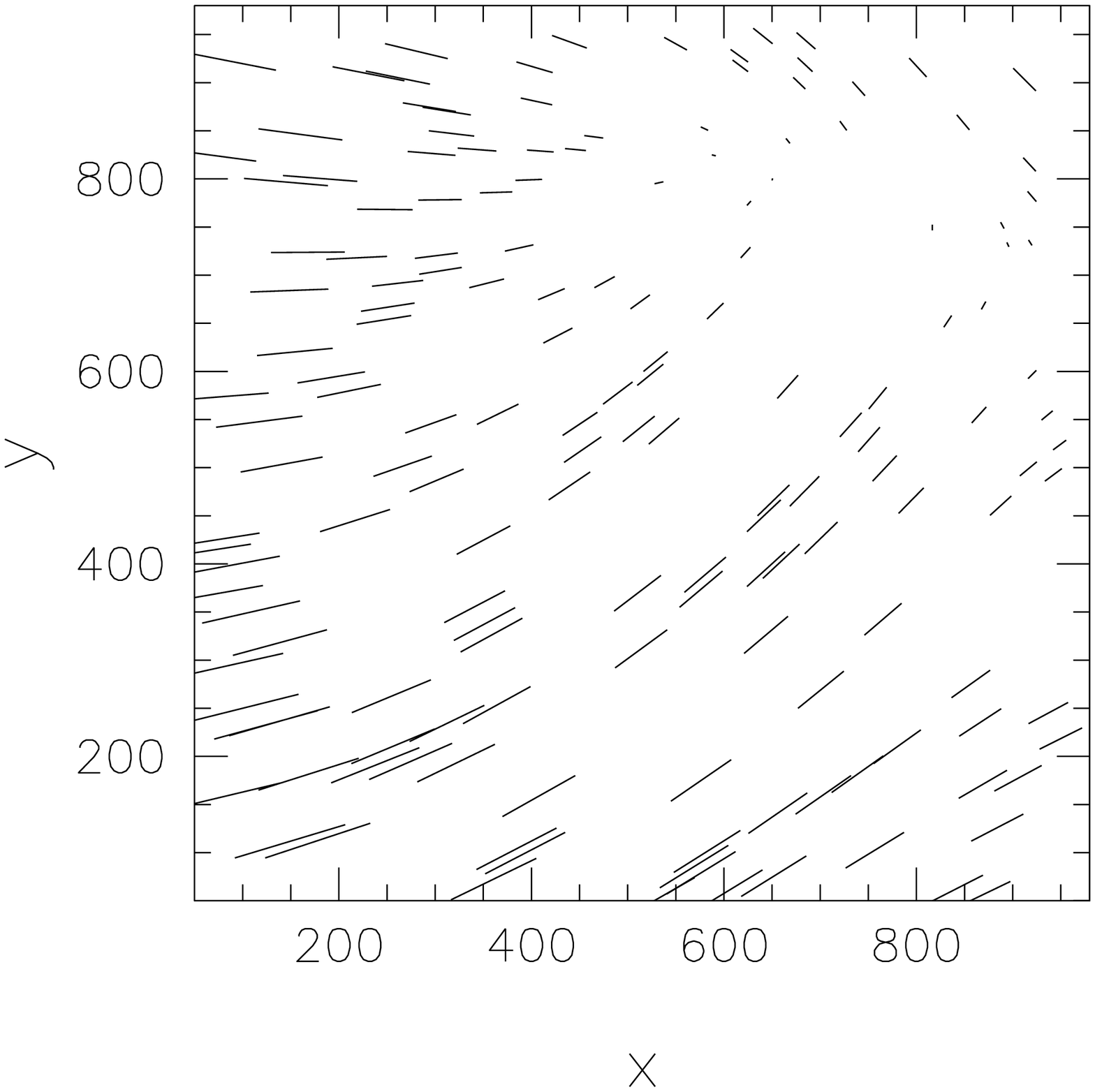}
 \hspace{0.2cm}
 \includegraphics[width=5cm]{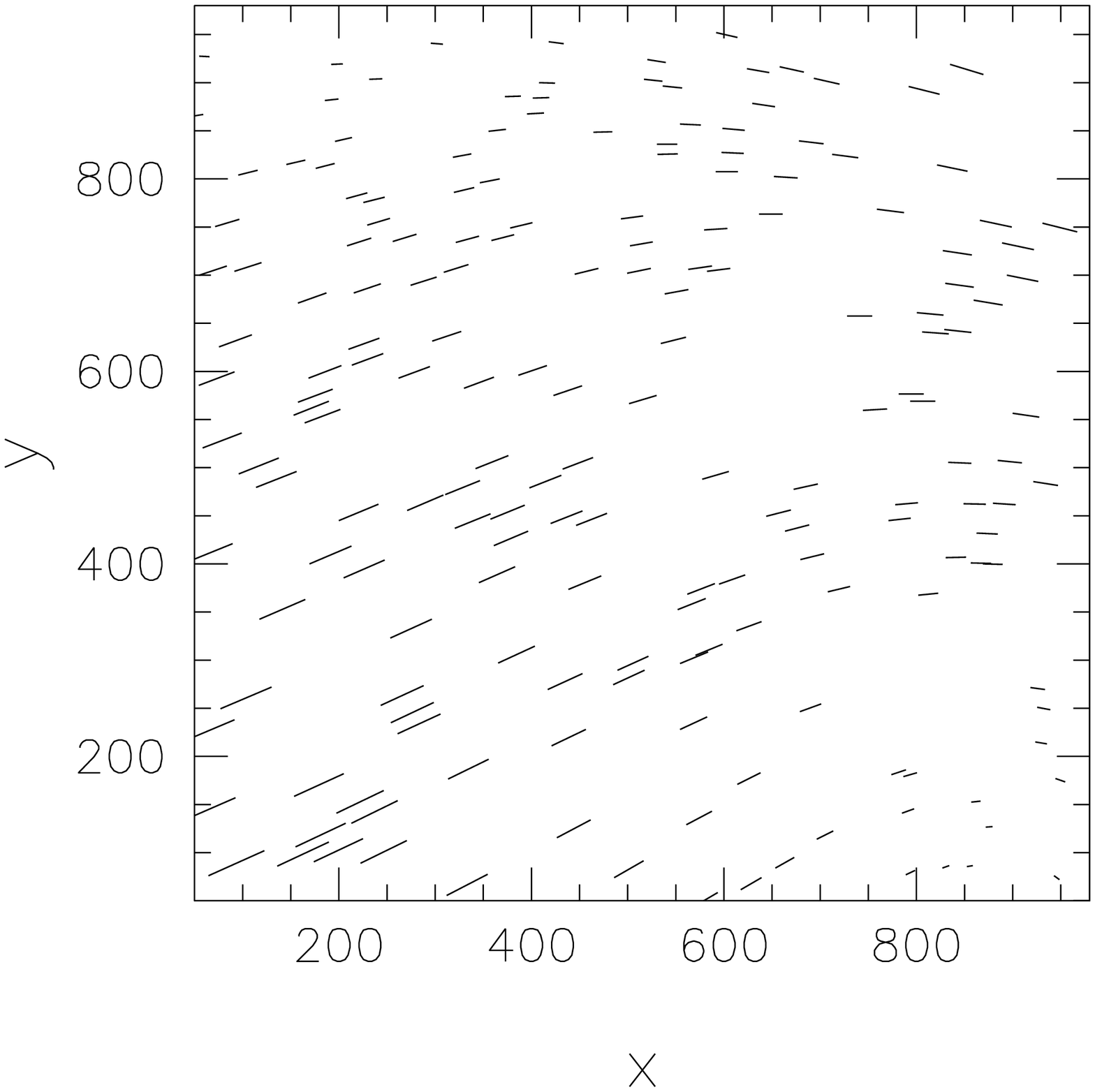}
\end{minipage}
\begin{minipage}{4.5cm}                      
 \caption{For one of the star fields the distribution of the
 ellipticities of stars are shown before (top left) and after (top
 right) correcting for PSF anisotropy. The middle left panel shows the
 spatial distribution of the ellipticities across the STIS field, the
 middle right the fitted second order polynomial at the star
 positions. The length of the sticks indicates the modulus of the
 ellipticity, the orientation gives the position angle. For individual
 exposures of this field, which were taken with a time difference of
 only 30 minutes we show the values of the polynomials for the
 anisotropy correction in the lower panels. One clearly sees the short
 timescale variations which are likely due to breathing of the
 telescope. }
 \label{fig:o3PSF}
\end{minipage}
\end{figure}

In addition to the variations from field to field, we find a spatial
variation of the PSF within individual fields, as shown in
Fig.~\ref{fig:o3PSF}, to which we fitted a second-order polynomial as
a function of position on the CCD. We also find some very short timescale
variations of the PSF (also shown in Fig.~\ref{fig:o3PSF}) which are
likely due to breathing of the telescope. 

\begin{figure}[!h]                             
\begin{minipage}{10.5cm}                        
 \includegraphics[width=10.5cm]{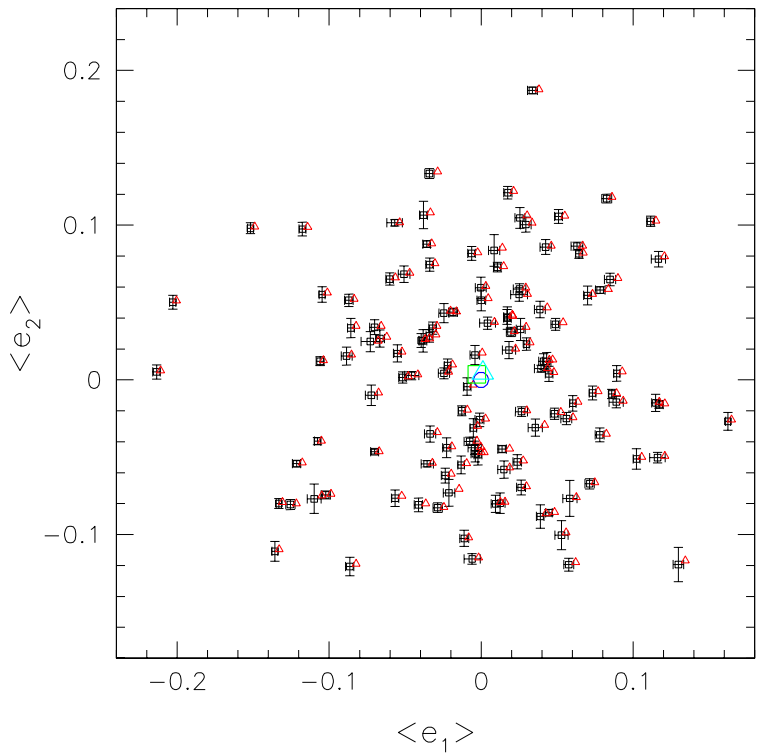}
\end{minipage}
\begin{minipage}{5.5cm}                        
 \caption{For 121 galaxy fields we plot the mean uncorrected
 ellipticity of galaxies (triangles) as well as the mean corrected
 ellipticity (squares). The error bars attached to the squares denote
 3 times the dispersion of the field-averaged corrected ellipticities
 when the different PSF model fits are used. The shift of the
 corrected mean ellipticities towards negative $e_1$ is expected
 from the behaviour of the stellar ellipticities plotted in
 Fig.~\ref{fig:aniexp}. The red triangle and cyan square in the center
 denote the mean over all galaxy fields of the uncorrected and
 corrected mean ellipticities, respectively; the size of the symbols
 show the $1\sigma$ errors on the mean. The circle shows the origin
 for reference. The mean is compatible with zero.}
 \label{fig:melli}
\end{minipage}
\end{figure}

In Fig.~\ref{fig:melli} we show that the PSF anisotropy corrections
are very small, they change the mean of the galaxy ellipticities on a
field only slightly. We also checked the effect of the anisotropy
correction on the cosmic shear result and find that it does not change
noticeably. 

\newpage
\clearpage

\mbox{}\\
\vspace{-2.7cm}
\mbox{}\\
\section{Results}

\begin{figure}[!t]                             
\begin{minipage}{5.5cm}                        
 \includegraphics[width=5.5cm]{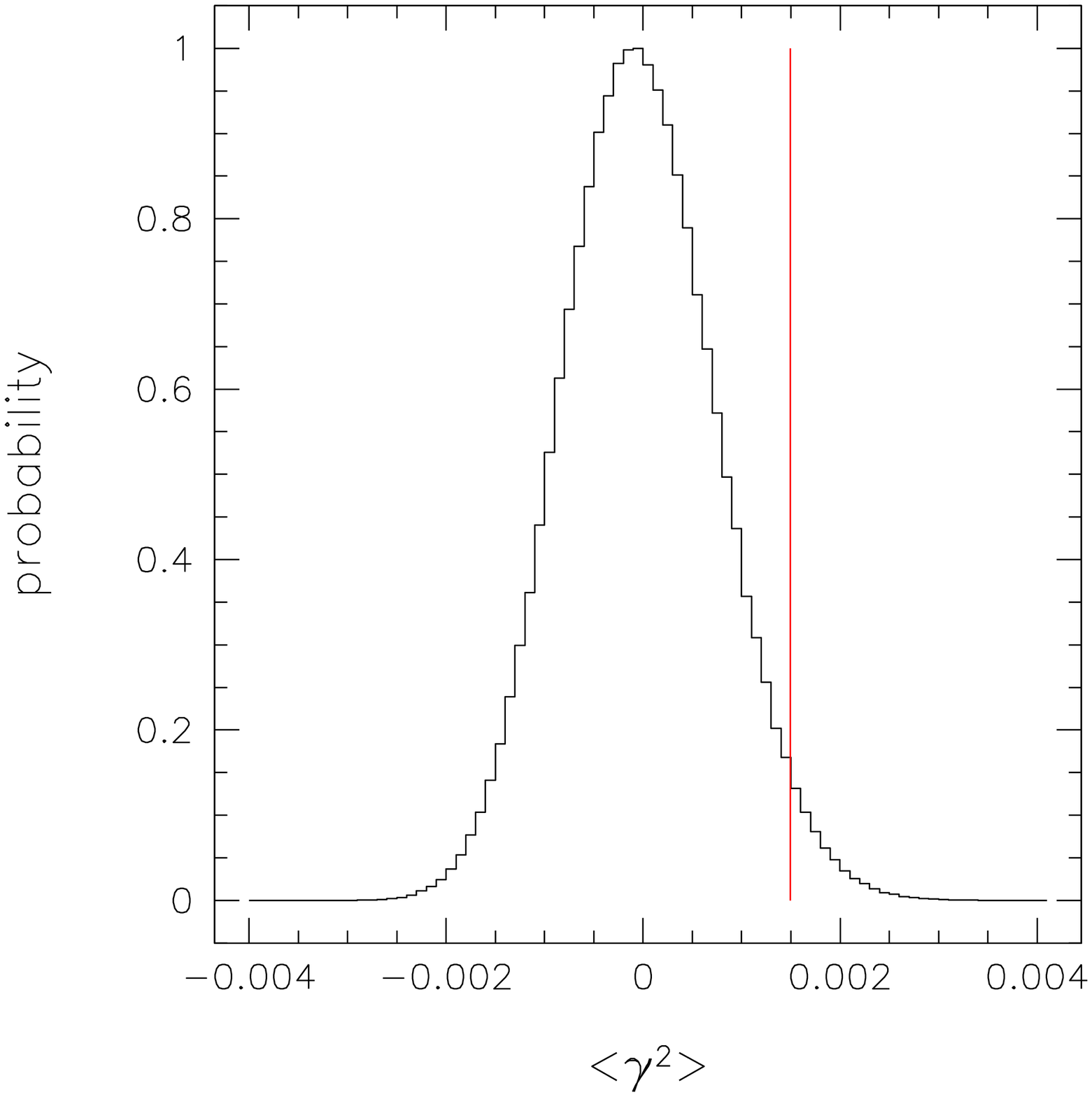}
\end{minipage}
\begin{minipage}{5.5cm}                        
 \includegraphics[width=5.5cm]{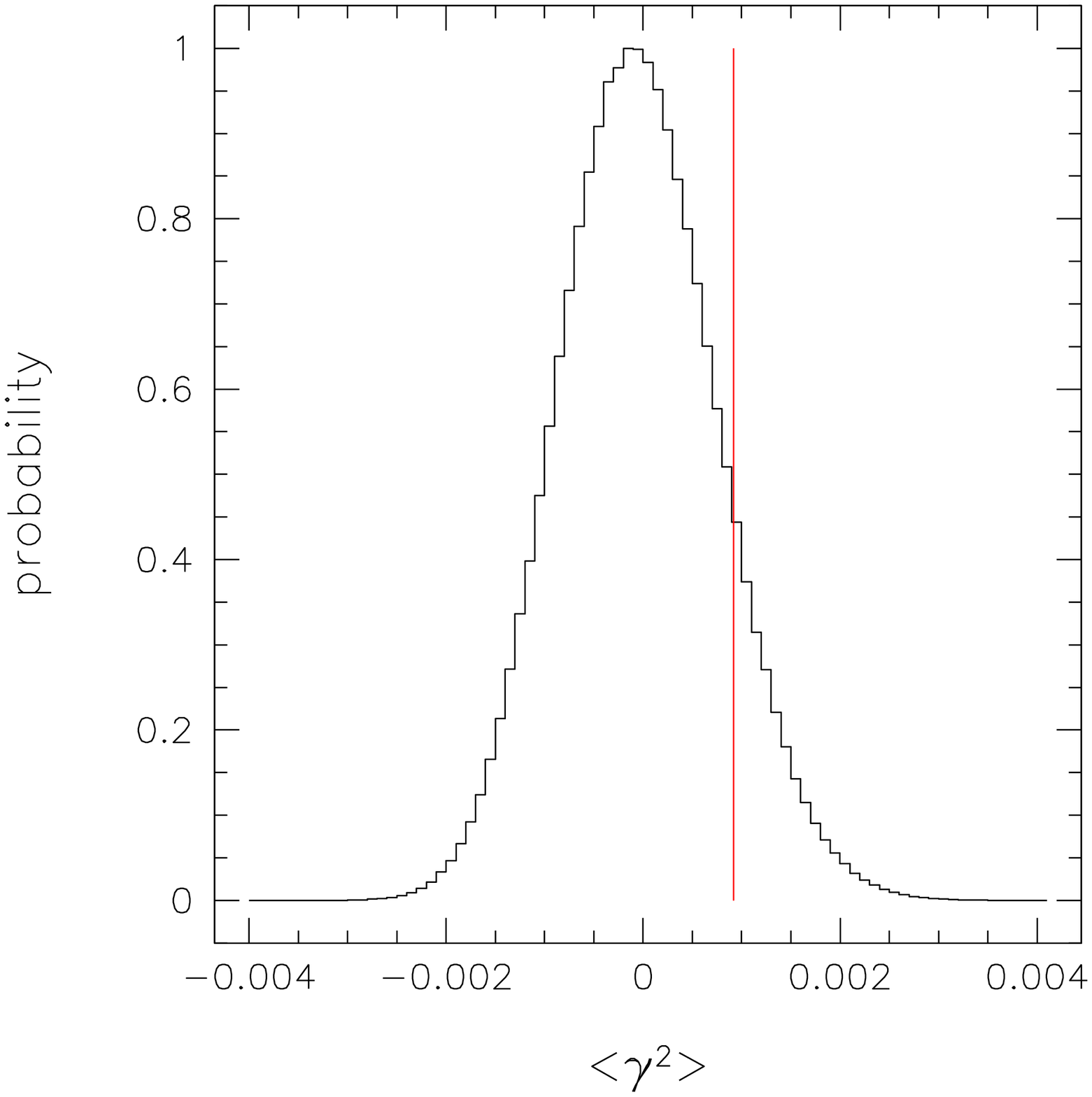}
\end{minipage}
\begin{minipage}{5.cm}                        
 \caption{Probability distribution of the cosmic shear dispersion estimator
 calculated from the data by randomizing the orientations of
 galaxies. The vertical line indicates the actually measured value;
 left is with,  right without weighting the galaxies
 depending on their noise level.}
\end{minipage}
 \label{fig:pdist}
\end{figure}

We analysed the ellipticities of galaxies in 121 galaxy fields, which
were corrected for PSF effects using 21 measured PSFs from low
galactic latitude fields (star fields). The quantity estimated was the
shear dispersion in each field. We apply a weighting to individual
galaxies according to the weighting scheme in Erben et al. (2001). 
By randomizing the orientations of the galaxy images, we
obtain probability distributions for our estimator in the absence of
shear, which are shown in Fig.~\ref{fig:pdist}, with and without
applying the weighting; this serves as an estimate of the significance
of our result. 

\begin{figure}[!h]                             
\begin{minipage}{8cm}                        
 \includegraphics[width=8cm]{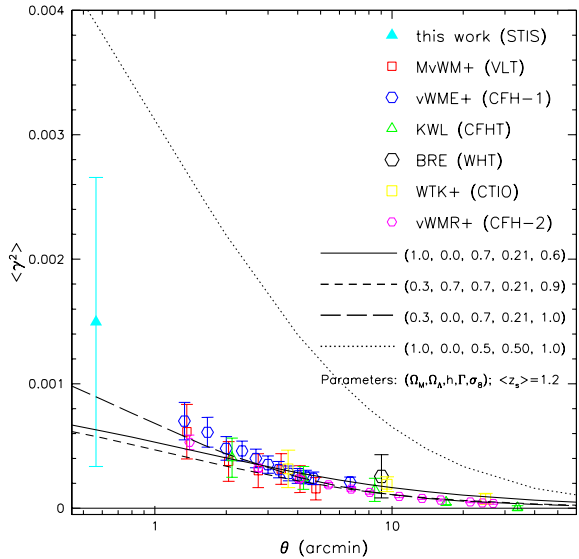}
\end{minipage}
\begin{minipage}{8cm}                        
 \caption{Comparison of our cosmic shear result (big triangle at
 $0.5^\prime$) with measurements at larger angular scales and with
 model predictions. The lines show the theoretical predictions if one uses
 different cosmological models, where we indicate $\Omega_m$,
 $\Omega_\Lambda$, $h$, $\Gamma$ and $\sigma_8$. The redshift
 distribution was taken from Brainard et al. (1996) with a mean source
 redshift of $\ave{z_\mathrm{s}}=1.2$. 
 Given the different galaxy selection implied by our STIS data, which
 may imply a substantially different redshift distribution from that
 obtained from ground-based observations, the apparent discrepancy may
 be due to a higher mean redshift. }
 \label{fig:cmp_cs}
\end{minipage}
\end{figure}

In Fig.~\ref{fig:cmp_cs} we compare our result for the cosmic shear
with weighting individual galaxies to
those obtained for larger scales by other groups. 
\mbox{}\\
\vspace{-0.25cm}
\mbox{}\\
{\small
\textit{Acknowledgements}
This work was supported by the TMR Network ``Gravitational Lensing: New
Constraints on Cosmology and the Distribution of Dark Matter'' of the
EC under contract No. ERBFMRX-CT97-0172  and by the German
Ministry for Science and Education (BMBF) through the DLR under the
project 50 OR106.
}
\newpage

\mbox{}\\
\vspace{-1.2cm}
\mbox{}\\
\textbf{\Large References}\\
\mbox{}\\
\vspace{-0.80cm}
\small
\mbox{}\\
 Bacon,~D., Refregier,~A., Ellis,~R.S., 2000, MNRAS, 318, 625\\[0.15cm]
 Brainerd,~T.G., Blandford,~R.D., Smail,~I., 1996, ApJ, 466, 623\\[0.15cm]
 Erben,~T., van Waerbeke,~L., Bertin,~E., Mellier,~Y., Schneider,~P.,
 2001, A\&A, 366, 717\\[0.15cm]
 Kaiser,~N., Wilson,~G., Luppino,~G.A., 2000, submitted to ApJ,
 astro-ph/0003338 \\[0.15cm]
 Maoli,~R.,  Mellier,~Y., van Waerbeke,~L., Schneider,~P., Jain,~B.,
 Bernardeau,~F., Erben,~T., Fort,~B., 2001, A\&A, 368, 766\\[0.15cm]
 Pirzkal,~N., Collodel,~L., Erben,~T., Fosbury,~R.A.E., Freudling,~W.,
 H\"{a}mmerle,~H., Jain,~B., Micol,~A., Miralles,~J.-M.,
 Schneider,~P. Seitz,~S., White,~S.D.M., 2001, A\&A in press,
 astro-ph/0102330 \\[0.15cm]
 van Waerbeke,~L.,  Mellier,~Y., Erben,~T., Cuillandre,~J.C.,
 Bernardeau,~F., Maoli,~R., Bertin,~E., Mc~Cracken,~H.J.,
 Le~F{\`e}vre,~O., Fort,~B., Dantel-Fort,~M., Jain,~B., Schneider,~P.,
 2000, A\&A, 358, 30 \\[0.15cm]
 van Waerbeke,~L.,  Mellier,~Y., Radovich,~M., Bertin,~E.,
 Dantel-Fort,~M., Mc~Cracken,~H.J., Le~F{\`e}vre,~O., Foucaud,~S.,
 Cuillandre,~J.C., Erben,~T., Jain,~B., Schneider,~P., Bernardeau,~F.,
 Fort,~B., 2001, A\&A in press, astro-ph/0101511\\[0.15cm]
 Wittman,~D.M., Tyson,~J.A., Kirkman,~D., Dell'Antonio,~I.,
 Bernstin,~G., 2000, Nature, 405, 143
\end{document}